\documentstyle[12pt,aaspp4]{article}

\begin{document}

\title{Observation of Gravitational Lensing in the Clumpy Universe} 


\author{Hideki ASADA\altaffilmark{1}}
\affil{ 
%
Yukawa Institute for Theoretical Physics \\
Kyoto University, Kyoto 606-01, Japan \\
{\it email: asada@yukawa.kyoto-u.ac.jp}
}
\altaffiltext{1}{Present address: Faculty of Science and Technology, 
Hirosaki University, Hirosaki 036, Japan; asada@phys.hirosaki-u.ac.jp}

\begin{abstract}
We discuss how inhomogeneities of the universe affect 
observations of the gravitational lensing; 
(1) the bending angle, (2) the lensing statistics and (3) the time delay. 
In order to take account of the inhomogeneities, the Dyer-Roeder distance 
is used, which includes a parameter representing the clumpiness of 
the matter along the line of sight. 
It is shown analytically that all three combinations of distances 
appearing in the above observations (1)-(3) are monotonic 
with respect to the clumpiness in general, 
for any given set of the density parameter, cosmological constant 
and redshifts of the lens and the source. 
Some implications of this result for the observation are presented; 
the clumpiness decreases both the bending angle and the lensing event rate, 
while it increases the time delay. 
We also discuss cosmological tests using the gravitational lensing 
in the clumpy universe. 
\end{abstract}



\section{Introduction}

It is one of the most important longstanding problems 
to determine the cosmological parameters (Weinberg 1972; Peebles 1993). 
There are some methods to determine the cosmological parameters 
by using the gravitational lenses (Refsdal 1964a, 1964b, 1966; 
Press and Gunn 1973; Blandford and Narayan 1986; 
Fukugita, Futamase and Kasai 1990; 
Fukugita, Futamase, Kasai and Turner 1992; 
Schneider, Ehlers and Falco 1992; hereafter SEF). 
Most of them can be classified into the following three typical observations: 
(1) the bending angle, (2) the lensing statistics and (3) the time delay. 
It is of great importance to clarify the relation between 
the observation in the realistic universe and the determination of 
the cosmological parameters. 
In particular, it has been discussed that inhomogeneities of 
the universe may affect the cosmological tests 
(Dyer and Roeder 1972, 1974; hereafter DR; 
Schneider and Weiss 1988a, 1988b; Linder 1988; Futamase and Sasaki 1989; 
Kasai, Futamase and Takahara 1990; Bartelmann and Schneider 1991; 
Watanabe, Sasaki and Tomita 1992; Sasaki 1993). 
In this paper, we use the Dyer-Roeder (DR) angular diameter distance 
in order to take account of the inhomogeneities (DR). 
Besides the three parameters (the Hubble constant, the density parameter and 
the cosmological constant), the DR distance contains an extra parameter 
representing the clumpiness of the matter along the line of sight. 
Two extreme cases can be represented by the DR distance; 
one is the distance in the Friedmann-Lemaitre-Robertson-Walker (FLRW) 
universe, the so-called filled beam, while the other is that of the 
so-called empty beam when the right ray propagates through the empty region. 
For comparison with the filled beam, the empty beam has been frequently 
and numerically used in the literature (For instance, 
Fukugita, Futamase and Kasai 1990; 
Fukugita, Futamase, Kasai and Turner 1992). 
However, it has not been clarified whether the observed quantities and/or 
the cosmological parameters obtained in the arbitrary case of 
the clumpiness parameter are bounded between those for the filled beam 
and the empty beam. 
Moreover, some cases of clumpiness parameter have been investigated 
numerically for fixed redshifts of the lens and the source 
(For instance, Alcock and Anderson 1985). 
However, the effect of the clumpiness on the observable depends on 
the redshifts of the lens and the source. 
Therefore, it is important to clarify how the observation of gravitational 
lensing depends on all the parameters (the density parameter, 
cosmological constant, clumpiness parameter 
and redshifts of the lens and the source), 
since the dependence seems complicated. 
Hence, we investigate analytically the arbitrary case of 
the clumpiness parameter, for any set of the density parameter, 
cosmological constant and redshifts of the lens and the source. 

This paper is organized as follows. 
Section 2 introduces three types of distance combinations 
appearing in the gravitational lensing (1)-(3). 
Section 3 shows the basic equations for the DR distance. 
Section 4 presents the proof that all these distance combinations 
are monotonic functions of the clumpiness parameter. 
We also clarify the effect of the clumpiness on the observations (1)-(3). 
In section 5, we discuss how the clumpiness affects the determination 
of the cosmological parameters. 
Conclusions are summarized in section 6.  

\section{Distance combinations appearing in gravitational lenses}

(1) bending angle

The lens equation is written as (SEF)  
\begin{equation}
\mbox{\boldmath $\beta$}=\mbox{\boldmath $\theta$}-{D_{LS} \over D_{OS}}
\mbox{\boldmath $\alpha$} . 
\label{angle}
\end{equation}
Here, {\boldmath $\beta$} and {\boldmath $\theta$} are 
the angular position vectors of the source and image, respectively, 
and {\boldmath $\alpha$} is the vector representing the deflection angle. 
$D_{LS}$ and $D_{OS}$ are the angular diameter distances from the lens to  
the source, and from the observer to the source, respectively. 
The effective bending angle $(D_{LS} / D_{OS})\mbox{\boldmath $\alpha$}$ 
appears when we discuss the observation concerning the angle such as 
the image separation and the location of the critical line 
(Blandford and Narayan 1986, SEF). 
Hence the ratio $D_{LS} / D_{OS}$ plays an important role in the discussion 
on the observation concerning the angle. 
It has been argued that, in calculating the bending angle, the density 
along the line of sight should be subtracted from the density of 
the lens object (Sasaki 1993). 
However, we assume that the density of the lens is much larger than 
that along the line of sight, so that the effect of the clumpiness 
on {\boldmath $\alpha$} can be ignored. 
Thus, we consider only the ratio $D_{LS} / D_{OS}$ in the following. 

(2) lensing statistics 

The differential probability of lensing events is (Press and Gunn 1973; SEF)
\begin{equation}
d\tau=\sigma n_L dl , 
\label{statistics}
\end{equation}
where 
$n_L$ is the number density of the lens, $dl$ is the physical length of the 
depth and $\sigma$ is the cross section proportional to 
$D_{OL}D_{LS} / D_{OS}$. 
Here, $D_{OL}$ is the angular diameter distance from the observer 
to the lens. 
Since $dl$ depends only on the cosmological parameters in the FLRW universe, 
we investigate the combination $D_{OL}D_{LS} / D_{OS}$ in order to 
take account of the clumpiness of the matter. 

(3) time delay 

The time delay between two images A and B is written as 
(Refsdal 1964b, SEF)
\begin{equation}
\Delta t_{AB}={1+z_L \over c}{D_{OL}D_{OS} \over D_{LS}}
\int^B_A d\mbox{\boldmath $\theta$}\cdot\Bigl(
{ \mbox{\boldmath $\alpha$}_A+\mbox{\boldmath $\alpha$}_B \over 2 }
-\mbox{\boldmath $\alpha$}(\mbox{\boldmath $\theta$}) \Bigr) , 
\label{delay}
\end{equation}
where {$\mbox{\boldmath $\alpha$}_A$} and {$\mbox{\boldmath $\alpha$}_B$} 
are the bending angles at the images A and B, respectively. 

\section{DR distances}

The DR angular diameter distance is determined by (DR, SEF, Sasaki 1992) 
\begin{equation}
{d^2 \over dw^2}D+{3 \over 2}(1+z)^5 \alpha \Omega D=0 , 
\label{dreq}
\end{equation}
where the parameter $\alpha$ represents the clumpiness of the matter 
along the light ray. 
In the FLRW universe, $\alpha$ is unity, while $\alpha$ vanishes 
when the light ray propagates through the empty space. 
Here $w$ is an affine parameter, which is assumed to be that 
in the FLRW universe, determined by 
\begin{equation}
{dz \over dw}=(1+z)^2 \sqrt{ \Omega z (1+z)^2 -\lambda z (2+z)+(1+z)^2 } , 
\label{affine}
\end{equation}
where $\Omega$ and $\lambda$ denote the density parameter and 
the cosmological constant, respectively. 
Since the coefficient of the last term of Eq.($\ref{dreq}$), 
$3\alpha\Omega/2$, comes from the Ricci focusing by the matter 
along the line of sight, the DR angular diameter distance is 
a decreasing function of $\alpha$ for a fixed redshift (DR, SEF). 
That is to say, 
\begin{equation}
D_{OL}(\alpha_1) > D_{OL}(\alpha_2) \quad\mbox{for}\quad 
\alpha_1 < \alpha_2 . 
\label{drtime1}
\end{equation}
In reality, the parameter $\alpha$ takes various values 
according to mass distribution. 
For instance, it can take a rather low value such as 0.5
in the clump model (Kasai, Futamase and Takahara, 1990; Linder, 1997). 

The DR equation $(\ref{dreq})$ must be solved under the boundary conditions, 
\begin{equation}
D(z_1,z_1)=0 , 
\label{bc1}
\end{equation}
and 
\begin{equation}
\left. {d \over dz_2}D(z_1,z_2)\right|_{z_2=z_1}=a(z_1){c \over H(z_1)} ,  
\label{bc2}
\end{equation}
where $a(z_1)$ and $H(z_1)$ denote the scale factor and the Hubble constant 
at $z_1$, respectively.

\section{Monotonic properties}

(1) $D_{LS}/D_{OS}$ 

It has been shown that the distance ratio $D_{LS}/D_{OS}$ satisfies 
(Asada 1997) 
\begin{equation}
{ D_{LS} \over D_{OS} }(\alpha_1) < 
{ D_{LS} \over D_{OS} }(\alpha_2) \quad\mbox{for}\quad \alpha_1 < \alpha_2 . 
\label{drbend1}
\end{equation}
This is proved as follows. 
For fixed $z_S$, $\Omega$ and $\lambda$, the ratio $D_{LS}/D_{OS}$ 
can be considered as a function of $z_L$, $X_{\alpha}(z_L)$. 
We define $Y_{\alpha}(z_L)$ as $D_{SL}/D_{OS}$, 
where $D_{SL}$ is the DR distance from the source to the lens. 
Owing to the reciprocity (SEF), we obtain 
\begin{equation}
Y_{\alpha}(z_L)={1+z_S \over 1+z_L}X_{\alpha}(z_L) . 
\label{reciprocity}
\end{equation}
Since $Y_{\alpha}$ depends on $z_L$ only through $D_{SL}$, 
it obeys the DR equation, 
\begin{equation}
{d^2 \over dw_L^2}Y_{\alpha}(z_L)+{3 \over 2}(1+z_L)^5 \alpha \Omega 
Y_{\alpha}(z_L)=0 , \label{raychaudhuri2}
\end{equation}
where $w_L$ is an affine parameter at the lens. 
We define the Wronskian as 
\begin{equation}
W(Y_{\alpha_1},Y_{\alpha_2})=\Bigl( Y_{\alpha_1}{d Y_{\alpha_2} \over 
d w_L}-Y_{\alpha_2}{d Y_{\alpha_1} \over d w_L} \Bigr) . 
\end{equation}
Using Eq.$(\ref{raychaudhuri2})$, we obtain 
\begin{equation}
{d \over d w_L}W(Y_{\alpha_1},Y_{\alpha_2}) < 0 
\quad\mbox{for}\quad \alpha_1 < \alpha_2 . 
\label{wronskian2}
\end{equation}
Since both $Y_{\alpha_1}$ and $Y_{\alpha_2}$ vanish at $z_L=z_S$, 
we obtain 
\begin{equation}
W(Y_{\alpha_1}(z_S),Y_{\alpha_2}(z_S))=0 . \label{wronskian3} 
\end{equation}
{}From Eqs.$(\ref{wronskian2})$ and $(\ref{wronskian3})$, we find 
\begin{equation}
W(Y_{\alpha_1},Y_{\alpha_2}) >0 , \label{wronskian4}
\end{equation}
where we used the fact that the affine parameter $w$ defined 
by Eq.$(\ref{affine})$ is an increasing function of $z$. 
Equation ($\ref{wronskian4}$) is rewritten as 
\begin{equation}
{d \over d w_L} \ln{Y_{\alpha_2} \over Y_{\alpha_1}} > 0 . 
\label{wronskian5}
\end{equation}
Since $Y_{\alpha}$ always becomes $1+z_S$ at the observer, we find 
\begin{equation}
\ln{Y_{\alpha_2}(z_L=0) \over Y_{\alpha_1}(z_L=0)}=0 . \label{wronskian6}
\end{equation}
{}From Eqs.$(\ref{wronskian5})$ and $(\ref{wronskian6})$, we obtain 
\begin{equation}
\ln{Y_{\alpha_2} \over Y_{\alpha_1}} >0 . \label{proof}
\end{equation}
This leads to 
\begin{equation}
{X_{\alpha_2} \over X_{\alpha_1}}>1 , \label{proof2}
\end{equation}
where we used  Eq.$(\ref{reciprocity})$. 
Thus, Eq.$(\ref{drbend1})$ is proved. 

It should be noted that Eq.($\ref{proof}$) holds even if one uses 
the opposite sign in the definition of the affine parameter 
in Eq.$(\ref{affine})$. 
{}From Eqs.($\ref{angle}$) and ($\ref{drbend1}$), the image separation 
as well as the effective bending angle increases with $\alpha$. 

(2) $D_{OL}D_{LS} / D_{OS}$ 

Next let us prove that $D_{OL}D_{LS} / D_{OS}$ increases monotonically 
with $\alpha$. 
We fix $\Omega$, $\lambda$, $z_L$ and $z_S$. 
Then it is crucial to notice that the distance from the lens 
to the source can be expressed in terms of the distance function 
from the observer, $D(z)$, as (Linder 1988)
\begin{equation}
D_{LS}={c \over H_0} (1+z_L) D_{OL}D_{OS} \int^{w_S}_{w_L} 
{dw \over D(z)^2} ,  
\label{dls}
\end{equation}
where $H_0$ is the Hubble constant at present. 
It is verified in a straightforward manner that Eq.$(\ref{dls})$ 
satisfies the DR equation $(\ref{dreq})$ with the boundary conditions 
$(\ref{bc1})$ and $(\ref{bc2})$. 
Equation $(\ref{dls})$ is rewritten as 
\begin{equation}
{D_{OL}D_{LS} \over D_{OS}}(\alpha)={c \over H_0}(1+z_L) D_{OL}\,^2 
\int^{w_S}_{w_L} {dw \over D(z)^2} . 
\label{drstat1} 
\end{equation}
The right hand side of this equation depends on $\alpha$ only through 
$D_{OL} / D(z)$. 
By the similar manner to the proof of Eq.$(\ref{drbend1})$, 
we obtain for $z_L<z<z_S$ 
\begin{equation}
{D_{OL} \over D(z)}(\alpha_1) < {D_{OL} \over D(z)}(\alpha_2) 
\quad\mbox{for}\quad \alpha_1 < \alpha_2 . 
\label{drstat2}
\end{equation}
By applying Eq.$(\ref{drstat2})$ to Eq.$(\ref{drstat1})$, 
we obtain 
\begin{equation}
{D_{OL}D_{LS} \over D_{OS}}(\alpha_1) < 
{D_{OL}D_{LS} \over D_{OS}}(\alpha_2) 
\quad\mbox{for}\quad \alpha_1 < \alpha_2 . 
\label{drstat3}
\end{equation}
Therefore, the gravitational lensing event rate increases with $\alpha$. 

(3) $D_{LS} / D_{OL}D_{OS}$

Finally, we investigate the combination of distances appearing 
in the time delay. 
By dividing Eq.$(\ref{drtime1})$ by Eq. $(\ref{drbend1})$, we obtain 
\begin{equation}
{D_{OL}D_{OS} \over D_{LS}}(\alpha_1) >
{D_{OL}D_{OS} \over D_{LS}}(\alpha_2) 
\quad\mbox{for}\quad \alpha_1 < \alpha_2 .
\label{drtime2}
\end{equation}
Thus, the time delay decreases with $\alpha$. 

Before closing this section, a remark is given: 
As shown above, the three types of combinations of distances 
are monotonic with respect to the clumpiness parameter. 
However, some of other combinations of distances are not monotonic 
functions of $\alpha$, though these combinations could not be necessarily 
relevant to physical problems. 
For instance, a combination $D_{LS} / \sqrt{c D_{OS} / H_0}$ 
is not a monotonic function of $\alpha$. 

\section{Implications for cosmological tests}

In this section, we consider the three types of the cosmological test 
which use combinations of distances appearing 
in the gravitational lensing. 
Let us fix the density parameter in order to discuss constraints on 
the cosmological constant. 

(1) $D_{LS}/D_{OS}$ 

The following relation holds 
\begin{equation}
{D_{LS} \over D_{OS}}(\lambda_1) < {D_{LS} \over D_{OS}}(\lambda_2) 
\quad\mbox{for} \quad \lambda_1 < \lambda_2 . 
\label{lambend1}
\end{equation}
This is proved as follows. 
Let us define 
\begin{equation}
X_{\lambda}(z_L)={D_{LS}(\alpha,\Omega,\lambda) \over 
D_{OS}(\alpha,\Omega,\lambda)} , 
\end{equation}
and 
\begin{equation}
Y_{\lambda}(z_L)={D_{SL}(\alpha,\Omega,\lambda) \over 
D_{OS}(\alpha,\Omega,\lambda)} . 
\end{equation}
By the reciprocity (SEF), we obtain 
\begin{equation}
Y_{\lambda}(z_L)={1+z_S \over 1+z_L}X_{\lambda}(z_L) . 
\label{reciprocity2}
\end{equation}
The ratio $Y_{\lambda}(z_L)$ satisfies 
\begin{equation}
{d^2 \over dw_L^2}Y_{\lambda}(z_L)+{3 \over 2}(1+z_L)^5 \alpha \Omega 
Y_{\alpha}(z_L)=0 . 
\label{raychaudhuri21}
\end{equation}
For $\lambda_i\,(i=1,2)$, the affine parameter $w_i$ satisfies 
\begin{equation}
{dz_L \over dw_i}=(1+z_L)^2 \sqrt{ \Omega z_L (1+z_L)^2 -\lambda_i 
z_L(2+z_L)+(1+z_L)^2 } . 
\label{affine21}
\end{equation}
We obtain 
\begin{equation}
{dz_L \over dw_1}>{dz_L \over dw_2} \quad\mbox{for}\quad 
\lambda_1 < \lambda_2 . 
\label{affine22}
\end{equation} 
We define the Wronskian as 
\begin{equation}
W(Y_{\lambda_1},Y_{\lambda_2})=\Bigl( Y_{\lambda_1}{d Y_{\lambda_2} \over 
d w_2}-Y_{\lambda_2}{d Y_{\lambda_1} \over d w_1} \Bigr) . 
\label{wronskian21}
\end{equation}
Using Eq.$(\ref{raychaudhuri21})$ and Eq.$(\ref{affine22})$, we obtain 
\begin{equation}
{d \over d z_L}W(Y_{\lambda_1},Y_{\lambda_2}) < 0 
\quad\mbox{for}\quad \lambda_1 < \lambda_2 . 
\label{wronskian22}
\end{equation}
Since $Y_{\lambda}$ always vanishes at $z_L=z_S$, we obtain 
\begin{equation}
W(Y_{\lambda_1}(z_S),Y_{\lambda_2}(z_S))=0 . 
\label{wronskian23} 
\end{equation}
{}From Eqs.$(\ref{wronskian22})$ and $(\ref{wronskian23})$, we find 
\begin{equation}
W(Y_{\lambda_1},Y_{\lambda_2}) >0 
\quad\mbox{for}\quad \lambda_1 < \lambda_2 . 
\label{wronskian24}
\end{equation}
This is rewritten as 
\begin{equation}
{d \over d z_L} \ln{Y_{\lambda_2} \over Y_{\lambda_1}} > 0 
\quad\mbox{for}\quad \lambda_1 < \lambda_2 ,  
\label{wronskian25}
\end{equation}
where we used Eq.($\ref{affine22}$). 
Since $Y_{\lambda}$ always becomes $1+z_S$ at the observer, we find 
\begin{equation}
\ln{Y_{\lambda_2}(z_L=0) \over Y_{\lambda_1}(z_L=0)}=0 . \label{wronskian26}
\end{equation}
{}From Eqs.$(\ref{wronskian25})$ and $(\ref{wronskian26})$, we obtain 
\begin{equation}
\ln{Y_{\lambda_2} \over Y_{\lambda_1}} >0 
\quad\mbox{for}\quad \lambda_1 < \lambda_2 . 
\label{proof21}
\end{equation}
This leads to 
\begin{equation}
{X_{\lambda_2} \over X_{\lambda_1}}>1 
\quad\mbox{for}\quad \lambda_1 < \lambda_2 
, \label{proof22}
\end{equation}
where we used  Eq.$(\ref{reciprocity2})$. 
Thus, Eq.$(\ref{lambend1})$ is proved. 

Together with Eq.$(\ref{drbend1})$, Eq.$(\ref{lambend1})$ means that, 
in the cosmological test using the bending angle, the cosmological constant 
estimated by the use of the distance formula in the FLRW universe is always 
{\it less} than that by the use of the DR distance $(0 \leq \alpha <1)$. 

(2) $D_{OL}D_{LS}/D_{OS}$ 

By multiplying Eq.$(\ref{lambend1})$ with 
\begin{equation}
D_{OL}(\lambda_1) < D_{OL}(\lambda_2) 
\quad\mbox{for}\quad \lambda_1 < \lambda_2 , 
\label{lamstat1}
\end{equation}
we obtain 
\begin{equation}
{D_{OL}D_{LS} \over D_{OS}}(\lambda_1) < 
{D_{OL}D_{LS} \over D_{OS}}(\lambda_2) 
\quad\mbox{for}\quad \lambda_1 < \lambda_2 . 
\label{lamstat2}
\end{equation}
Equation $(\ref{lamstat1})$ can be proved, for instance 
in the following manner: 
The DR distance is written as the integral equation 
(Schneider and Weiss 1988a, Linder 1988) 
\begin{equation}
D(z;\alpha)=D(z;\alpha=1)+{3 \over 2}{c \over H_0} (1-\alpha) \Omega
\int^{z}_0 dy \left.{dw \over dz}\right|_{z=y} (1+y)^4 
D(y,z;\alpha=1)D(y;\alpha) . 
\label{DRlam}
\end{equation}
This is rewritten as (Schneider and Weiss 1988a, Linder 1988)
\begin{equation}
D(z;\alpha)=D(z;\alpha=1)+\sum^{\infty}_{i=1} \left[ {3 \over 2} 
{c \over H_0} (1-\alpha) \Omega \right]^i \int^z_0 dy K_i(y,z) 
D(y;\alpha=1) , 
\label{DRlam2}
\end{equation}
where $K_i(y,z)$ is defined as 
\begin{equation}
K_1(x,y)=\left.{dw \over dz}\right|_{z=x} (1+x)^4 D(x,y;\alpha=1) 
\label{K1}
\end{equation}
and 
\begin{equation}
K_{i+1}(x,y)=\int^y_x dz K_1(x,z)K_i(z,y) . 
\label{Ki}
\end{equation}
{}From Eqs.$(\ref{affine22})$, $(\ref{K1})$ and $(\ref{Ki})$, 
it is shown that for $x<y$ 
\begin{equation}
K_i(x,y;\lambda_1) < K_i(x,y;\lambda_2) 
\quad \mbox{for} \quad \lambda_1 < \lambda_2 , 
\label{Kilam}
\end{equation}
where we used the following relation in the FLRW universe as 
\begin{equation}
D(x,y;\alpha=1,\lambda_1) < D(x,y;\alpha=1,\lambda_2) 
\quad \mbox{for} \quad \lambda_1 < \lambda_2 . 
\label{FLRWlam} 
\end{equation}
Using Eqs.$(\ref{DRlam2})$, $(\ref{Kilam})$ and $(\ref{FLRWlam})$, 
and the positivity of $K_i$, we find Eq.$(\ref{lamstat1})$. 

{}From Eqs.$(\ref{drstat3})$ and $(\ref{lamstat2})$, it is found that, 
in the cosmological test using the lensing events rate, 
the cosmological constant is always {\it underestimated} 
by the use of the distance formula in the FLRW universe.

(3) $D_{LS}/D_{OL}D_{OS}$ 

We consider the combination $D_{LS} / D_{OL}D_{OS}$. 
Since the time delay is measured and the lens object is observed, 
$D_{OL}D_{OS} / D_{LS}$ can be determined from Eq.($\ref{delay}$). 
On the other hand, when we denote the dimensionless distance between 
$z_1$ and $z_2$ as $d_{12}=H_0 D_{12} / c$, which does not depend on 
the Hubble constant, we obtain 
\begin{equation}
{D_{OL}D_{OS} \over D_{LS}}={c \over H_0} {d_{OL}d_{OS} \over d_{LS}} . 
\label{delay2}
\end{equation}
Then, Eq.($\ref{drtime2}$) becomes 
\begin{equation}
{d_{OL}d_{OS} \over d_{LS}}(\alpha_1) > {d_{OL}d_{OS} \over d_{LS}} 
(\alpha_2) \quad\mbox{for}\quad \alpha_1 < \alpha_2 . 
\label{drtime3}
\end{equation}
Thus, from Eqs.($\ref{delay2}$) and ($\ref{drtime3}$), it is found that 
$H_0$ estimated by using the DR distance decreases with $\alpha$. 
Thus, the Hubble constant can be bounded from below when we have little 
knowledge on the clumpiness of the universe. 
The lower bound is given by the use of the distance 
in the FLRW universe. 
On the other hand, the combination $D_{LS} / D_{OL}D_{OS}$ 
is {\it not} a monotonic function of the cosmological constant. 
Therefore, the relation between the clumpiness of the universe and 
the cosmological constant is not simple, since it depends on 
many parameters $(z_L, z_S, \Omega, \Lambda, \alpha)$. 

It should be noted that even the assumption of the spatially flat 
universe $(\Omega+\lambda=1)$ does not change the above implications 
for the three types of cosmological tests, since 
the cosmological constant affects the DR distance formula 
only through the relation between $z$ and $w$, Eq.$(\ref{affine})$. 

\section{Conclusion}

We have investigated the effect of the clumpiness of the matter along the 
line of sight on the three types of observations on 
(1) the bending angle, (2) the lensing statistics and (3) the time delay. 
First, it has been shown analytically that the combinations of 
distances which appear in the gravitational lensing (1)-(3) 
are monotonic with respect to the clumpiness parameter $\alpha$ 
when the DR distance is used. 
This property presents us with the following implications 
for the observation: 
In the clumpy universe approximated by the DR distance, the bending angle is 
{\it smaller}, the lensing events occur {\it less frequently} and 
the time delay is {\it longer} than in the FLRW universe. 
In the cosmological tests using (1) the bending angle and (2) 
the lensing statistics, the use of the DR distance  always leads to 
the {\it overestimate} of the cosmological constant. 
However, the same is not true of the time delay, since 
the combination of DR distances in the time delay is not monotonic with 
respect to the cosmological constant. 
Whether the effect of the clumpiness parameter enhance the effect of the 
cosmological parameter or not depends on other parameters. 
Rather, the primary cosmological use of the time delay is the physical 
estimation of the Hubble constant 
(Refsdal 1964b, Blandford and Kundic 1996). 
It has been found that the use of the DR distance never lowers 
the estimate of the Hubble constant from the time delay. 

We have taken the clumpiness parameter $\alpha$ as a constant 
along the line of sight. 
However, as an extension of the DR distance, $\alpha$ can be considered 
phenomenologically as a function of the redshift in order to 
take account of the growth of inhomogeneities of the universe (Linder 1988). 
However, in the above consideration, particularly in the proof of 
the monotonic properties, it has never been used that $\alpha$ is constant. 
Hence, all the monotonic properties and the implications for cosmological 
tests remain unchanged for variable $\alpha(z)$. 
That is to say, when $\alpha_1(z) < \alpha_2(z)$ is always satisfied 
for $0<z<z_S$, all we must to do is to replace parameters $\alpha_1$ and 
$\alpha_2$ with functions $\alpha_1(z)$ and $\alpha_2(z)$ in 
Eqs.$(\ref{drbend1})$, $(\ref{drstat3})$ and $(\ref{drtime2})$.  
In particular, when $\alpha(z)$ is always less than unity on the 
way from the source to the observer, both of the three combinations of 
distances appearing in (1) and (2) are less, and the combination 
in (3) are larger than those in the FLRW universe. 
Then, the decrease in the bending angle and the lensing event rate, 
and the increase in the time delay hold even for a generalized DR 
distance with variable $\alpha(z)$. 

The DR distance is useful for theoretical and conceptual studies 
on the lensing in the clumpy universe. 
However, the DR distance seems too simple to describe the realistic 
universe where the density fluctuation is of stochastic nature. 
In fact, there are many lines of sight on which $\alpha$ may be deviating 
appreciably from unity (Kasai, Futamase and Takahara 1990; Linder 1997). 
The conclusion obtained by the use of the DR distance encourages 
us to make a statistical discussion on the observation 
of gravitational lensing in the clumpy universe. 
For such a discussion, it is necessary to investigate 
the light propagation in the realistic universe numerically 
(Schneider and Weiss 1988a, 1988b; Bartelman and Schneider 1991; 
Watanabe, Sasaki and Tomita 1992) 
and perturbatively (For instance, Seljak 1994; Bar-Kana 1996). 


\acknowledgements

The author would like to thank M. Sasaki for useful 
discussion and helpful comments on the earlier version
of the manuscript. 
He also would like to thank M. Kasai and T. Tanaka 
for fruitful discussion. 
He would like to thank S. Ikeuchi and T. Nakamura 
for continuous encouragement. 
This work is supported in part by Soryushi Shogakukai. 



\end{document}